\PassOptionsToPackage{numbers,sort&compress}{natbib}
\newif\ifrevtexavailable
\IfFileExists{revtex4-2.cls}{\revtexavailabletrue
  \documentclass[reprint,groupedaddress,nofootinbib,prl,aps,floatfix]{revtex4-2}
}{\revtexavailablefalse
  \documentclass[10pt,twocolumn]{article}
  \usepackage[a4paper,margin=1.8cm]{geometry}
}
\usepackage[T1]{fontenc}
\usepackage[utf8]{inputenc}
\usepackage[english]{babel}
\usepackage{amsmath,amssymb,bm}
\usepackage{graphicx,booktabs,array}
\usepackage{natbib}
\usepackage[colorlinks=true,linkcolor=black,citecolor=blue,urlcolor=blue]{hyperref}
\newcommand{\ec}{\mathrm{EC}}
\newcommand{\betap}{\beta^+}
\newcommand{\Ree}{\operatorname{Re}}
\newcommand{\epsT}{\epsilon_T}
\newcommand{\epsS}{\epsilon_S}

\newcommand{\abstracttext}{We introduce the ratio of electron capture to positron decay rates of unstable nuclei as a probe for exotic tensor currents in the electroweak sector. We calculate nuclear structure corrections and newly derived radiative corrections and obtain competitive limits on exotic tensor couplings using historically available data to find $\varepsilon_T = (1.06 \pm 0.76)\times 10^{-3}$ which is competitive with collider searches and the global $\beta$-decay database. In doing so, we resolve a 50-year-old tension in the strongly hindered $^{22}$Na transition by retaining higher-order operators, obtaining a consistent description of its different weak-decay observables. Finally, we discuss the potential of this method to significantly improve tensor constraints using modern cryogenic quantum sensors.}
\begin{document}
\title{Competitive constraints on exotic tensor currents from $\ec/\beta^+$ ratios}
\ifrevtexavailable
\author{Victor Dumenil}
\author{Leendert Hayen}
\email[Corresponding author: ]{hayen@lpccaen.in2p3.fr}
\affiliation{LPC Caen, ENSICAEN, Universit\'e de Caen, CNRS/IN2P3, Caen, France}
\else
\author{Victor Dumenil and Leendert Hayen\\
\small LPC Caen, ENSICAEN, Universit\'e de Caen, CNRS/IN2P3, Caen, France\\
\small\texttt{hayen@lpccaen.in2p3.fr}}
\fi
\date{9 October 2026}
\ifrevtexavailable
\begin{abstract}\abstracttext\end{abstract}
\maketitle
\else
\maketitle
\begin{abstract}\abstracttext\end{abstract}
\fi

\textit{Introduction.---} Exotic scalar or tensor currents in electroweak processes show up generically in many Beyond the Standard Model extensions, but remain unobserved despite several decades of intense research \cite{Falkowski2020}. Neutron and nuclear $\beta$ decays, in particular, probe such interactions in ways competitive and complementary with collider searches \cite{Gonzalez-Alonso2018}. During the last decade a strong focus has been on direct $\beta$ spectroscopy, motivating substantial ab initio theory work \cite{Glick-Magid2022, King2022} as well as the development of new techniques \cite{Byron2023}. Even so, many precise spectroscopic measurements are hampered by several systematic effects and progress has been slower than anticipated. Alternative channels with enhanced sensitivity and different experimental uncertainties are therefore particularly valuable.

We may write the appearance of exotic currents below the electroweak scale using the low-energy effective field theory (LEFT) Lagrangian \cite{Gonzalez-Alonso2018,Hayen2024},
\begin{align}
\mathcal L={}&-\frac{G_F\widetilde V_{ud}}{\sqrt2}\biggl\{
\bar e\gamma_\mu(1-\gamma^5)\nu_e\,
\bar u\gamma^\mu[1-(1-2\epsilon_R)\gamma^5]d\nonumber\\
&+\epsS\,\bar e(1-\gamma^5)\nu_e\,\bar u d\nonumber\\
&+\epsT\,\bar e\sigma_{\mu\nu}(1-\gamma^5)\nu_e\,
\bar u\sigma^{\mu\nu}(1-\gamma^5)d+\ldots\biggr\}
+\mathrm{h.c.},
\label{eq:LEFT}
\end{align}
where $G_F$ is Fermi's constant and $\widetilde V_{ud}=V_{ud}(1+\epsilon_L+\epsilon_R)$ is the up--down quark mixing matrix element modified by exotic interactions. At lowest order, the exotic interactions arise from dimension-six operators probing new physics at a scale $\Lambda_{\mathrm{BSM}}$ via $\epsilon\sim(m_W/\Lambda_{\mathrm{BSM}})^2$, with $m_W$ the $W$ boson mass. Low energy searches using neutron and nuclear $\beta$ decay probe such currents via the appearance of the Fierz interference term
\begin{align}
\mathcal B&=\frac{1}{1+\rho^2}\Ree\left\{\frac{g_S}{g_V}\epsS-4\rho^2\frac{g_T}{g_A}\epsT\right\},\nonumber\\
b_{\ec}&=+2\gamma_{\ec}\mathcal B,
\qquad b_{\betap}=-2\gamma_\beta\mathcal B,\label{eq:bF}
\end{align}
where we neglected $\epsilon^2$ effects, and which modifies each decay rate via $d\Gamma_X=d\Gamma_{X,\mathrm{SM}}[1+b_X(m_e/E_e)]$. Here $g_A=1.263(10)$, $g_S=1.085(114)$ and $g_T=0.981(21)$ are positive nucleon isovector charges \cite{Aoki2021}, $\rho=g_AM_{GT}/g_VM_F$ is the ratio of axial-to-vector matrix elements, and $\gamma_X=\sqrt{1-(\alpha Z_X)^2}$. For a pure Gamow--Teller (GT) transition, the GT fraction is unity and the scalar contribution vanishes.

Exotic scalar searches are strongly constrained using measured lifetimes of superallowed $0^+\to0^+$ vector decays, resulting in $\epsS=0.0001(10)$ \cite{Hardy2020, Falkowski2020}. Tensor-like interactions, on the other hand, are probed through axial transitions. The 2021 global analysis of beta-decay experiments finds $\epsT^{\beta}=0.000125(325)$ in our convention \cite{Falkowski2020}. Nuclear structure and other higher-order effects appear at the percent level in conventional observables, and theoretical uncertainties must be evaluated on a case-by-case basis.

We may construct an observable that is largely free of these drawbacks and presents a substantially increased sensitivity to exotic currents. In the presence of the latter, we may write the relative decay rates of the electron capture and $\beta^+$ processes for the same transition as
\begin{align}
R\equiv\frac{\lambda_{\ec}}{\lambda_{\betap}}
&=R_{\mathrm{SM}}\,
\frac{1+b_{\ec}\langle m_e/E_e\rangle_{\ec}}
{1+b_{\betap}\langle m_e/E_e\rangle_\beta},\label{eq:ratio_EC_beta}\\
R_{\mathrm{SM}}&=\frac{\sum_i n_i C_i f_i^{\ec}(1+\Delta_{R,i}^{\ec})}
{\overline C_\beta f_\beta(1+\Delta_R^\beta)}.\label{eq:RSM}
\end{align}
Here the sum runs over the different atomic orbitals with occupation $n_i$, bound-electron total energy $E_i=m_e-B_i$, binding energy $B_i$, and capture phase-space factors $f_i^{\ec}$. The EC average uses the same shell weights as the rate. Nuclear structure effects are contained in the shape factors $C_{\ec,\beta}$, and both processes contain radiative corrections denoted $\Delta_R^{\ec,\beta}$. For $\beta^+$ decay,
\begin{equation}
\overline C_\beta=
\frac{\int_{m_e}^{E_0}dE\,pE(E_0-E)^2F(Z_d,E)K(Z_d,E)C(E)}
{\int_{m_e}^{E_0}dE\,pE(E_0-E)^2F(Z_d,E)K(Z_d,E)},
\label{eq:Cmean}
\end{equation}
where $E_0$ is the spectral endpoint, $F$ the positron Fermi function, $Z_d$ the daughter charge, and $K$ contains the remaining spectral corrections \cite{Hayen2018,Hayen2019a}.

Equation~\eqref{eq:ratio_EC_beta} contains a number of distinct advantages: $(i)$ the sensitivity to the Fierz interference is greatly enhanced due to the opposite signs. With $E_i\approx m_e$, the two contributions add as $\langle m_e/E_e\rangle_{\ec}+\langle m_e/E_e\rangle_\beta$, providing a gain of a factor of several relative to the integrated positron rate alone; $(ii)$ the leading matrix element precipitating the reaction is shared between the EC and $\beta^+$ modes and drops out, while cancellation between higher-order form factors substantially suppresses the remaining nuclear structure sensitivity, typically to the per-mille level; $(iii)$ experimentally, the determination of Eq.~\eqref{eq:ratio_EC_beta} is a much simpler \textit{counting} experiment, significantly reducing energy spectroscopy artifacts plaguing traditional methods.

Many experimental determinations of the EC to $\beta^+$ ratio were performed already several decades ago for mostly for other purposes at the (few-)percent level \cite{Bambynek1977}, but stalled in part due to limited theoretical agreement. As we will mention below, however, studies on $^{22}$Na had been motivated by the search for a non-zero Fierz term in the 1960's, but was found to be particularly difficult to interpret \cite{Firestone1975a}. Since then, computational and conceptual improvements in atomic and nuclear calculations have changed the landscape. Here, we perform dedicated calculations of nuclear matrix elements to high orders using the nuclear shell model for precisely measured transitions as well as promising candidates for a future research program. We will show how this historically available data already provide tensor constraints comparable to global beta-decay analyses, with considerable scope for improvement, once nuclear structure and radiative corrections are properly taken into account.

\textit{Nuclear structure.---}We calculate nuclear structure effects using the Behrens--B\"uhring formalism \cite{Behrens1971, Behrens1982} which performs a multipole decomposition of hadronic and leptonic currents in the $\beta$ decay interaction. Subsequently, the shape factor $C$ (see Eq. (\ref{eq:ratio_EC_beta})) may be obtained as a multi-parameter expansion in the momentum transfer, $q$, the nuclear charge, $Z$, and the electron energy $E_e$. The tree-level matrix element cancels out in the ratio and a leading-order approximation of the remaining nuclear structure effects was given in Ref. \cite{Bambynek1977}
\begin{equation}
    \delta_{NS} \equiv \frac{C_i^{EC}}{\overline{C}_\beta} -1 \sim \frac{A_0+\rho^2A_1}{1+\rho^{2}}
    \label{eq:delta_NS}
\end{equation}
where $A_i$ depends on matrix elements of rank $i$, which to first order are part of
\begin{equation}
    \bm{r}\times\bm{p}, \quad\bm{r}^2, \quad\bm{\sigma} \times (\bm{r}\times \bm{p}),\quad \gamma_5\bm{r},
\end{equation}
with $\bm{r}(\bm{p})$ the nucleon position (momentum) and $\bm{\sigma}$ the Pauli matrix. Due to the similarity of the processes, contributions in $A_i$ partially cancel. For mirror transitions, the dominant weak magnetism form factor can additionally be constrained by magnetic dipole moments through the conserved vector current relation \cite{Severijns2023}.

\begin{ruledtabular}
    \begin{table}[ht]
    \caption{Nuclear structure corrections for a selection of isotopes with precise data (top) and low-mass mirror isotopes that form promising candidates for future study (bottom). $\delta_{NS}^{LO}$ refers to the leading-order approximation in Eq. (\ref{eq:delta_NS}) whereas $\delta_{NS}^{tot}$ is obtained after proper convergence of the power series. Uncertainties are due to the interactions used in the nuclear shell model as detailed in the text.}
        \centering
        \begin{tabular}{c|cccccc}
            Isotope & $\rho_{exp}$ & $\rho_{th}$ & $A_0 (\%)$ & $A_1(\%)$ & $\delta_{NS}^{LO}(\%)$ & $\delta_{NS}^{tot}(\%)$\\
            \hline
            $^{11}$C & -0.7544(8) & -0.73(4) & 0.81 & -0.06 & 0.25 & 0.30(1) \\
            $^{22}$Na & - & - & - & -1.28 & -1.28 & -1.40(7) \\
            $^{57}$Ni & - & - & - & 0.09 & 0.09 & 0.03(1) \\
            $^{58}$Co & - & 2.50 & -0.21 & -0.47 & -0.44 & -0.85(17) \\
            $^{65}$Zn & - & - & - & -0.10 & -0.10 & 0.37(7) \\
            \hline
            $^{13}$N & -0.560(1) & -0.53(5) & 0.40 & -0.08 & 0.04 & 0.02(2) \\
            $^{15}$O & 0.630(2) & 0.54(2) & 0.41 & -0.11 & 0.04 & 0.01(3) \\
            $^{17}$F & 1.296(1) & 1.15(1) & 0.99 & -0.17 & 0.56 & 0.64(1) \\
            $^{18}$F & - & - & - & 0.54 & 0.54 & 0.66(1) \\
            $^{19}$Ne & -1.602(1) & -1.63(1) & 0.82 & -0.22 & 0.53 & 0.63(4) \\
        \end{tabular}
        \label{tab:delta_NS}
    \end{table}
\end{ruledtabular}

The leptonic contributions entering the shape factor are computed by expanding the lepton wave functions in powers of $m_eR$, $E_eR$ and $\alpha Z$. This expansion renders the nuclear matrix elements, also referred to as form factor coefficients, independent of the lepton momenta. On the hadronic side, the form factors are constructed within the impulse approximation from one-body transition densities (OBTDs) computed using KSHELL code \cite{Shimizu2013}. We outline the procedure in more detail in the end matter. The shell model code provides the OBTDs required to evaluate the matrix elements of the operators entering the Behrens–Bühring expansion. Within the impulse approximation, these matrix elements are obtained by combining the OBTDs with the corresponding single-particle matrix elements. They are then used to construct the nuclear form-factor coefficients entering the shape factor $C$. The latter combines the nuclear matrix elements with the leptonic radial wave-function expansion and determines the energy-dependent decay rates for EC and $\beta^+$ emission. The detailed nuclear calculations will be presented in a companion study \cite{DumenilInPreparation}, with a particular focus on the resolution of the anomalous $^{22}$Na decay.

For each transition, we first evaluate the shell model matrix elements entering the Behrens–Bühring expansion and construct the corresponding shape factors for EC and $\beta^+$ decay. We then calculate $\delta_{NS}$ from the ratio of the EC shape factor to the phase space averaged $\beta^+$ shape factor, according to Eq.~(\ref{eq:delta_NS}). We evaluate the leading-order expression separately and compare it with the result obtained using the higher-order contributions retained in our calculation (see Table~\ref{tab:delta_NS}). This procedure quantifies the impact of the nuclear matrix elements beyond the leading-order approximation on the predicted EC/$\beta^+$ ratio.

To distinguish the higher-order contributions, we separate the nuclear structure correction $\delta^{tot}_{NS}$ into contributions constrained by experimental information and contributions obtained from nuclear structure calculations,
\begin{equation}
\delta^{tot}_{NS} = \delta_{NS}^{exp} + \delta_{NS}^{theo} .
\end{equation}
Here, $\delta_{NS}^{exp}$ denotes the contribution evaluated using experimentally determined matrix elements for the leading allowed transitions (Fermi, Gamow-Teller and weak magnetism matrix elements)
as input, while $\delta_{NS}^{theo}$ is computed using higher-order NMEs with the Behrens-Bühring formalism. This separation allows us to distinguish the effects constrained by experimental data from those that depend more directly on the nuclear structure model. To compute the OBTDs, we used the following interactions: \textit{CKPOT} (for ${^{11}C}$, ${^{13}N}$ and ${^{15}O}$) \cite{Cohen1965}, \textit{USDB} (for ${^{17-18}F}$, ${^{19}Ne}$ and ${^{22}Na}$) \cite{Brown2006}, \textit{YSOX} (for ${^{11}C}$, ${^{13}N}$, ${^{15}O}$, ${^{17-18}F}$ and ${^{19}Ne}$) , \textit{HO} (for ${^{57}Ni}$), \textit{CA48MH1} (for ${^{58}Co}$) \cite{Hjorth-Jensen1995} and \textit{JUN45} (for ${^{65}Zn}$) \cite{Honma2005}. 
No truncation of the model space or modifications to the single-particle energies were made when computing the OBTDs.


To assess the reliability of the shell model calculations, we calculated the theoretical axial-to-vector matrix element ratio $\rho_{th}$ for the mirror transitions (see Table~\ref{tab:delta_NS}) and the theoretical partial half-lives and compared them with the experimental values. The calculated half-lives are generally reproduced within $20$–$30\%$, providing a consistency check of the nuclear wave functions and the dominant nuclear matrix elements. However, this agreement does not directly quantify the uncertainty associated with the higher-order matrix elements entering $\delta_{NS}$. In the absence of a systematic uncertainty estimate based on the spread of $\delta_{NS}$ values obtained with different interactions, we adopt $0.20\times|\delta_{NS}^{\mathrm{theo}}|$ as a phenomenological estimate of the associated nuclear structure uncertainty.


\textit{The $^{22}$Na puzzle.---}The longstanding disagreement associated with $^{22}$Na extends beyond the capture ratio to its beta spectrum shape, beta--gamma anisotropy and longitudinal positron polarization \cite{Firestone1978,Bowers1999,Skalsey1985, Triambak2017}. Its large $\log ft\simeq8$ reflects a severe cancellation in the leading GT strength, enhancing the relative contribution of higher-order operators. Consequently, its nuclear correction is the largest of the transitions considered here. The usual truncation to weak magnetism and induced tensor terms \cite{Holstein1971a,Holstein1977,Firestone1978} is insufficient in this unusually hindered transition. With a careful nuclear structure calculation and the additional operators retained, we find consistent agreement with all four experimental observables using the same nuclear inputs. In particular, the $\beta$-$\gamma$ anisotropy \cite{Triambak2017} is very sensitive to both $K=2,3$ order matrix elements. The historical anomaly can thus be understood without compromising the broader advantages of the channel. Additionally, we perform a new analysis of the various $^{22}$Na $EC/\beta^+$ measurements. Instead of the unweighted average in Ref. \cite{Mougeot2018}, we perform a statistical analysis of discrepant data and we use instead $(EC/\beta^+)_{22} = 0.1107(5)$, where the uncertainty has been increased by a factor 2.5. We report on all of this in more detail in a companion article \cite{DumenilInPreparation}.

\textit{Atomic and radiative corrections.---}The atomic EC rate must be evaluated at a precision commensurate with the experimental ratio. The most precise calculations of the $EC/\beta^+$ were performed by Mougeot \cite{Mougeot2018, Mougeot2019} within a metrology context. The approach is based on the methods of Bahcall \cite{Bahcall1962a, Bahcall1963a} and Vatai \cite{Vatai1970}. Shell-specific capture fraction calculations are particularly sensitive to shake-up and shake-off, as well as exchange and overlap corrections. In the current study we follow the results of Ref. \cite{Mougeot2019}. It has been noted already in Bahcall's original approach \cite{Bahcall1963a}, and reiterated in Ref. \cite{Bambynek1977}, that the total EC rate, on the other hand, is much less sensitive to either shaking, overlap or exchange corrections by invoking the closure relationship. In a separate work, we update the many-electron calculation using this approach the calculated moments of the EC amplitude \cite{HayenAtomicInPreparation}.

Radiative corrections at $\mathcal{O}(\alpha)$ for electron capture, which corresponds to real photon emission and wave function renormalization effects, were originally calculated by Holstein \cite{Holstein1979a} and revisited by Kurylov \textit{et al.} \cite{Kurylov2003}. At $\mathcal O(\alpha^2 Z)$, however, corrections are available only for the $\beta^+$ fraction \cite{Jaus1970, Sirlin1987b} and have recently regained interest \cite{Crosas2026}. At the per-mille level, however, higher-order radiative corrections to EC are not \textit{a priori} negligible. What's more, the radiative correction to the EC/$\beta^+$ ratio must be matched consistently to the atomic Hamiltonian, since vacuum-polarization effects may already be included through the bound-state potential. We have derived new $\mathcal{O}(\alpha^2Z)$ radiative corrections to the $EC/\beta^+$ ratio in an approach consistent with the atomic calculations, and crossing the relevant parts of the the continuum result to the bound-state channel. In particular, the common short-distance logarithm cancels in the ratio, leaving a finite regular contribution together with the difference of the EC and $\beta^+$ vacuum-polarization terms, and we find
\begin{align}
\delta_{\mathrm{EC}/\beta^+}^{(\alpha^2 Z)} &= Z\alpha^2 \left[ \frac{37}{8} -4\ln 2 +\left\langle F(\beta)\right\rangle_{\beta^+} \right] \nonumber \\
&+\delta_U^{\mathrm{EC}} -\left\langle\delta_U^{\beta^+}\right\rangle
\label{eq:alpha2Z_matching}
\end{align}
where $F(\beta)$ is the finite-$\beta$ continuum function of Ref. \cite{Crosas2026} and $\delta_U$ denotes the Uehling contribution evaluated with the same nuclear charge distribution.
We find that the ratio with the order $\alpha$ radiative corrections are considerable larger than $\alpha Z$, which can be understood from the Dirac-Coulomb wave function convergence in the bound state case. A detailed derivation is given in Ref.~\cite{HayenRCInPreparation}. For $^{22}$Na, Eq. \ref{eq:alpha2Z_matching} contributes a $-0.11\%$ decrease in the ratio compared to the $0.47\%$ order $\alpha$ result, while for $^{58}$Co the $\alpha^2Z$ contributions is practically equal but opposite to the $\mathcal{O}(\alpha)$ result.

\begin{figure}[t]
\centering
\includegraphics[width=0.48\textwidth]{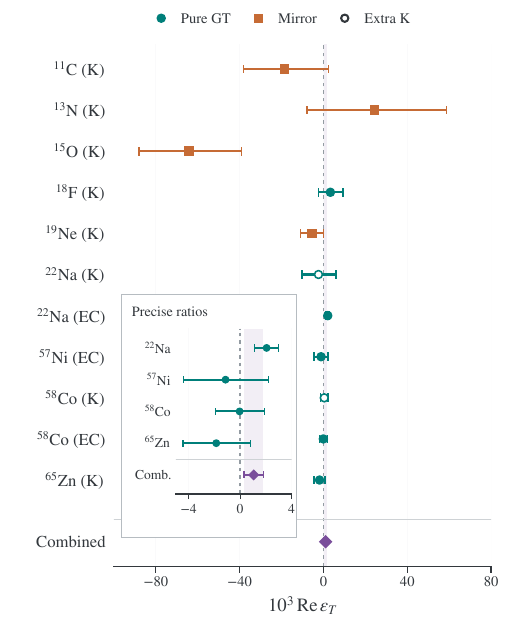}
\caption{Overview of individual EC$/\beta^+$ and $K/\beta^+$ ratios interpreted as constraints on $\varepsilon_T$. All results use the normalization of Eq.~\eqref{eq:LEFT} and one-standard-deviation errors.}
\label{fig:individual_constraints}
\end{figure}

\textit{Constraints and prospects.---}We combine nine measured ratios, using K capture for $^{11}$C, $^{13}$N, $^{15}$O, $^{18}$F, $^{19}$Ne and $^{65}$Zn and inclusive EC for $^{22}$Na, $^{57}$Ni and $^{58}$Co \cite{Bambynek1977,Mougeot2019}, shown in Fig. \ref{fig:individual_constraints}. 
We fit Eq.~\eqref{eq:ratio_EC_beta} directly, combining experimental, atomic and nuclear uncertainties. Scalar $\varepsilon_S$ and lattice-charge uncertainties are common inputs; at the present precision their inclusion changes the tensor uncertainty by less than $0.1\%$.

Combining the nine independent EC/$\beta^+$ measurements, we obtain
\begin{equation}
    \epsilon_T=0.00105(48)_{\rm exp}(59)_{\rm atom}(4)_{\rm NS}(4)_{\rm RC}
\end{equation}
corresponding to a total uncertainty of $7.6\times 10^{-4}$ in our tensor-operator convention. The result is consistent with the Standard Model at 1.4$\sigma$. The uncertainty is dominated by the atomic reference rates and experimental ratios, while the nuclear corrections and retained radiative-correction uncertainties are subdominant, identifying improved atomic calculations and precision counting measurements as the principal routes to stronger tensor constraints. In particular, the experimental uncertainty is inflated by a factor 2.5 due to the poor agreement between different $^{22}$Na results between the 1950's and 1980's, highlighting the potential for modern measurements.

\begin{figure}[t]
\centering
\includegraphics[width=0.48\textwidth]{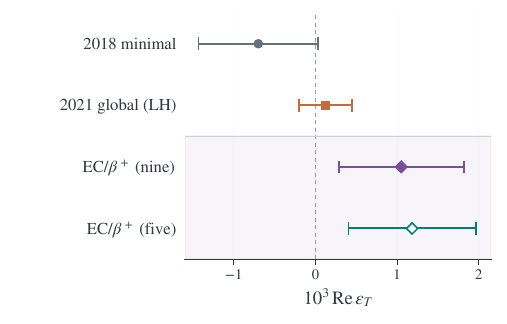}
\caption{Comparison of EC to positron tensor constraints with global beta-decay analyses. All results use the normalization of Eq.~\eqref{eq:LEFT} and one-standard-deviation errors. The 2018 minimal analysis is that of Ref. \cite{Gonzalez-Alonso2018}, and the 2021 global results from the purely left-handed analysis of Ref. \cite{Falkowski2020}.}
\label{fig:constraints}
\end{figure}

The comparison with global $\beta$-decay fits is shown in Fig. \ref{fig:constraints}. The comparison in Fig.~\ref{fig:constraints} uses the same operator normalization and confidence level: the 2018 minimal fit gives $\epsT=-0.0007\pm0.00073$ \cite{Gonzalez-Alonso2018}, while the 2021 global fit gives $\epsT=0.000125\pm0.000325$ for left-handed neutrinos \cite{Falkowski2020}. Historically available EC to positron data therefore already yield competitive constraints on exotic tensor currents.

Modern cryogenic quantum sensors provide a direct route to improved counting precision \cite{Hayen2024}. Superconducting tunnel junctions, for example, measure the recoil and atomic relaxation energy with eV-scale resolution, providing a means to distinguish capture and positron-emission events \cite{Fretwell2020, Friedrich2021, Kim2024}. 
Selecting unhindered transitions minimizes the remaining nuclear sensitivity, while improved atomic and radiative calculations reduce the theoretical uncertainty. These developments make EC to positron ratios particularly promising for substantial progress in Fierz searches.

\textit{Conclusions.---}Electron capture to positron ratios provide a particularly promising channel for Fierz constraints. The opposite signs of the interference enhance the sensitivity, the leading nuclear matrix element drops out, and the measurement is a counting experiment with reduced dependence on detector response. Existing data already provide competitive tensor constraints. The difficult nuclear structure of $^{22}$Na accounts for its exceptional place in the historical literature: retaining the enhanced higher-order operators brings its different weak-decay observables into consistent agreement. Together with more precise atomic rates and matched radiative corrections, the new experimental possibilities offered by direct nuclear recoil energy detectors open a path to substantially stronger limits on exotic currents.

\section*{Acknowledgements}
This research is supported by the French National Agency for Research, Agence National de la Recherche. This work has received support from the French government, managed by the National Research Agency (ANR), under the France 2030 program as part as CaeSAR project, reference "ANR-23-EXES-0001”, and the Normandy Region.

\bibliography{library}

\clearpage
\section*{End Matter}
\textit{Conventions.---}We use $\hbar=c=1$, $\sigma^{\mu\nu}=i[\gamma^\mu,\gamma^\nu]/2$ and $\gamma^5=i\gamma^0\gamma^1\gamma^2\gamma^3$. Real Wilson coefficients and scalar/tensor charges are quoted at $\mu=2$~GeV in $\overline{\mathrm{MS}}$, assuming left-handed neutrinos. The leptonic factor $1-\gamma^5$ is explicit in Eq.~\eqref{eq:LEFT}. The tensor coefficient follows Ref.~\cite{Gonzalez-Alonso2018}, without a factor $1/4$ multiplying the operator. Ref.~\cite{Falkowski2020} uses $\epsT^{\mathrm{FGN}}/4$ for the same operator, hence
\begin{equation}
\epsT^{\mathrm{FGN}}=4\epsT,\qquad
\frac{C_T^+}{C_A^+}=-4\frac{g_T}{g_A}\epsT,
\end{equation}
at leading matching order, with $C_A^+\propto-g_A$ and positive $g_A,g_T$. For pure GT, $f_{GT}=1$, $f_F=0$, rather than $\rho=1$, and
\begin{equation}
b_{\ec}=-8\gamma_{\ec}\frac{g_T}{g_A}\Ree\epsT,
\quad b_{\betap}=+8\gamma_\beta\frac{g_T}{g_A}\Ree\epsT.
\end{equation}
Thus positive $\epsT$ lowers EC to positron ratios. We use the parent charge in $\gamma_{\ec}$ and the daughter charge in $\gamma_\beta$, in the leading allowed Coulomb treatment. The captured-electron energy is $m_e-B_i$, not its binding energy. Inclusive EC factors are shell weighted. The rates retain interference linear in exotic amplitudes; their rational ratio does not include the omitted quadratic contributions.

To compute the shape factor $C$, we use the following expression, written as a sum over the rank $K$ and the angular momenta of the emitted electron $k_e$ and neutrino $k_\nu$:
\begin{align}
C(E_e) = \sum_{k_{e}, k_{\nu}, K \ge 0} \lambda_{k_{e}} \Big [ M_{K}(k_{e}, k_{\nu})^{2} + m_{K}(k_{e}, k_{\nu})^{2} \nonumber \\
- \frac{2\gamma_{k_{e}}\mu_{k_e}}{k_{e}E_e} M_{K}(k_{e}, k_{\nu})m_{K}(k_{e}, k_{\nu})\Big ] 
\label{eq:Shape factor}
\end{align}
This term represents a dynamical correction containing all the nuclear-structure information of the transition in terms of Nuclear Matrix Elements (NMEs), where $\lambda_{k_e}$ and $\mu_{k_e}$ are Coulomb functions defined as
\begin{equation}
\lambda_{k_e} = \frac{\alpha^2_{-k_e} + \alpha^2_{+k_e}}{\alpha^2_{-1} + \alpha^2_{+1}} , ,
\end{equation}
\begin{equation}
\mu_{k_e} = \frac{\alpha^2_{-k_e} - \alpha^2_{+k_e}}{\alpha^2_{-k_e} + \alpha^2_{+k_e}} \frac{k_eW_e}{\gamma_{k_e}} , .
\end{equation}
Here, $\alpha_{\pm k}$ denotes the Coulomb amplitudes, obtained by solving the Dirac equation for an electron moving in a uniform nuclear potential. The quantities $M_K(k_e, k_\nu)$ and $m_K(k_e, k_\nu)$ contain the convolution of the leptonic wave functions and the nuclear-structure information encoded in the form factors $F_{KLs}(q^2)$. The integers $k_e$ and $k_\nu$ are defined as $k_i = |\kappa_i|$, where $\kappa_i$ is the angular momentum quantum number of particle $i$.

The summation in Eq.~(\ref{eq:Shape factor}) can be separated into two cases: (1) $k_e + k_\nu = K + 1$ and (2) $k_e + k_\nu = K + 2$, with the constraint $|J_i - J_f| \leq K \leq J_i + J_f$.

The leptonic contributions entering the coefficients $M_K(k_e, k_\nu)$ and $m_K(k_e, k_\nu)$ are computed by expanding the lepton wave functions in powers of $(m_eR)$, $(E_eR)$, and $(\alpha Z)$. This expansion avoids the explicit calculation of overlap integrals between nuclear and leptonic wave functions, and renders the nuclear matrix elements-also referred to as form factor coefficients-independent of the lepton momenta.

On the hadronic side, the form factors appearing in $M_K(k_e, k_\nu)$ and $m_K(k_e, k_\nu)$ coefficients in Eq.~(\ref{eq:Shape factor}) can be expanded in powers of $(qR)$ for $q \ll 1$ as
\begin{equation}
F_{KLs}(q^2) = \sum_n \frac{(-1)^n(2L+1)!!}{(2n)!!(2L+2n+1)!!}(qR)^{2n}F^{(n)}_{KLs} , .
\end{equation}
In order to simplify the nuclear decay process, which involves solving a many-body system governed by complex interactions, the impulse approximation is employed for the decaying nucleus. Within this approximation, the decaying nucleon is treated independently from the remaining nucleons, which act only as spectators, while meson-exchange currents and other many-body effects are neglected. After introducing this approximation, the relation between the form factors and the NMEs can be written as
\begin{align}
{^{V/A}F}_{KLs}(q^2)&={^{V/A}F^{(N)}_{KLs}(k_e, m, n, \rho)} \nonumber \\
&\propto \sum_{\alpha, \beta} {^{V/A}\mathcal{M}^{(N),(\alpha, \beta)}_{KLs}(k_e, m, n, \rho)} \nonumber \\
&\times \langle f||[a_{\alpha}^{\dagger}a_{\beta}]_{K}||i\rangle ,
\label{eq:nmes to ff relation}
\end{align}
where $\langle f||[a_{\alpha}^{\dagger}a_{\beta}]_{K}||i \rangle$ is the one-body transition density (OBTD) between the nuclear states $\alpha$ and $\beta$, and is computed using a shell model code (here KSHELL).

The subscripts ($V/A$) in Eq.~(\ref{eq:nmes to ff relation}) correspond to the Vector and Axial-vector components of the weak interaction. In most applications of $\beta$-decay theory, only the leading-order NMEs are considered, typically the Fermi matrix element $M_F(^{V}F^{0}_{000})$, the Gamow-Teller matrix element $M_{GT}(^{A}F^{0}_{101})$, and the weak magnetism contribution $(^{V}F^{0}_{111})$. However, for high-precision low-energy measurements, it is necessary to extend the calculations to higher orders.

\textit{Confidence levels and correlated inputs.---}The 2018 minimal-fit uncertainty combines its fit and hadronic errors at 90\% confidence, $\sqrt{0.0012^2+0.0001^2}$, and is divided by $1.64485$ for comparison at $1\sigma$. The 2021 left-handed result is divided by four to match Eq.~\eqref{eq:LEFT}. The scalar input $\epsS=0.0001(10)$ and the lattice charges are shared across transitions. K and inclusive EC determinations for the same isotope are not counted independently.

\end{document}